# Investigation of spin-phonon coupling and local magnetic properties in magnetoelectric Fe$_2$TeO$_6$

P. Pal[1], Shalini Badola[2], P. K. Biswas[3], Ranjana R. Das[4] Surajit Saha[2], S. D. Kaushik[5], Parasmani Rajput[6], P. N. Vishwakarma[1], A. K. Singh[1]

[1]Department of Physics and Astronomy, National Institute of Technology, Rourkela-769008, India

[2]Department of Physics, Indian Institute of Science Education and Research, Bhopal-462066, India

[3]ISIS Facility, Science and Technology Facilities Council Rutherford Appleton Laboratory, Harwell Science and Innovation Campus, Chilton, Didcot OX11 0QX, UK

[4]Institut Néel - CNRS, 25 Avenue des Martyrs - BP 166, 38042 Grenoble cedex 9, France

[5]UGC-DAE Consortium for Scientific Research, Bhabha Atomic Research Centre Mumbai-400085, India

[6]Atomic & Molecular Physics Division, Bhabha Atomic Research Center, Trombay, Mumbai-400085, India

**Abstract:**

Spin-phonon coupling originated from spin-lattice correlation depends upon different exchange interactions in transition metal oxides containing 3d magnetic ions. Spin-lattice coupling can influence the coupling mechanism in magnetoelectric material. To understand the spin-lattice correlation in inverse trirutile Fe$_2$TeO$_6$ (FTO), magnetic properties and phonon spectra are studied. Signature of short-range magnetic correlation induced by 5/2-5/2 dimeric interaction and magnetic anomaly at 150 K is observed apart from the familiar sharp transition ($T_N$~210K) corresponding to long-range order by magnetization and heat capacity measurements. The magnetic transitions and the spin dynamics are further locally probed by muon spin resonance (μSR) measurement in both zero fields (ZF) and longitudinal field (LF) mode. Three dynamically distinct temperature regimes; (i) $T > T_N$, (ii) $T_N > T > 150$ K, and (iii) $T < 150$ K, are observed. A change in the spin dynamics is realized at 150 K by μSR, though previous studies suggest long-range antiferromagnetic order. The renormalization of phonon frequencies observed in Raman spectra below 210 K suggests the existence of spin-phonon coupling in the material. The coupling strength is quantified as in the range 0.1-1.2 cm$^{-1}$ following the mean-field and two-spin cluster approximations. We propose that the spin-phonon coupling mediated by the Fe-O2-Fe interbilayer exchange play a significant role in magnetoelectric ME coupling observed in the material.

## 1. Introduction:

The cross-coupling of spin, lattice, charge, and orbital degrees of freedom breed out large cornucopias of materials having dramatically complex electronic ground states with numerous intriguing and novel physical phenomena like high-$T_C$ superconductivity, colossal magnetoresistivity, magnetoelectric (ME) multiferroicity, *etc.*[1–5]. Research interest on the coupling between lattice and spin degrees of freedom, known as spin-phonon coupling (SPC), is continuously reviving from the perspective of ME multiferroic [5–8]. The success of SPC to explain thermal Hall Effect in some multiferroic drives more attention to understand the SPC in multiferroic material [9]. SPC can be employed to determine the spin relaxation time in spintronic applications like quantum computation [10]. The origin of SPC in transition metal oxides containing 3d and 4d ions depends upon the different exchange interactions present in the system[11]. Hence, the study of SPC, *i.e.,* the interdependency of spin and lattice in the magnetically ordered state is very important to understand magnetoelectricity and other coexisting phenomena in a multiferroic material.

The inverse trirutile ($P4_2/mnm$) Fe$_2$TeO$_6$ (FTO) is a collinear antiferromagnetic (AFM) material possessing ME coupling below its AFM transition $T_N$ at 210 K[12–14]. Magnetic correlation above $T_N$ is suspected in FTO as the presence of a broad anomaly above the sharp transition in magnetization at $T_N$[15,16]. A sharp rise in magnetization is also observed below 50 K, as reported earlier. Based on macroscopic magnetic behavior, different distinct regions of interest can be identified in the material; (i) broad transition around above T$_N$~210 K, (ii) sharp fall below $T_N$, and (iii) a sharp rise below 50 K. The spin dynamics of the system in all these three regions are essentially important to understand the ME coupling in the material. Some isostructural compounds of the inverse trirutile compounds are reported to possess short-range dimeric interaction well above $T_N$[17,18]. In this regard, muon spin relaxation (μSR) is an efficient technique that can be carried out in true zero fields (ZF). ZF-μSR is an extremely sensitive probe detecting tiny internal fields on the order of 0.1G[19]. It is also a very sensitive probe to study the magnetic fluctuation in the range of $10^4$ to $10^{12}$ Hz.

Recently, FTO the material is observed to show ME coupling and ferroelectricity above $T_N$ in both bulk and nano polycrystalline form [20,21]. ME coupling in this material was suspected of inducing by two ion mechanisms though the role of spin-lattice correlation to ME coupling was not explored[14]. The presence of ME coupling above $T_N$ was proposed to be originated due to the presence of short-range magnetic order present in the system. In one of our very recent works, the ferroelectricity and ME coupling are observed to be correlated with local $d^5$ off-centering of $Fe^{3+}$ polyhedra [22]. Locally distorted geometries with shared (edge, corner, or face) binuclear units play a crucial role in determining spin, hybridization with ligand, and magnetic ground state in transition metal compounds[23]. The $Fe^{3+}$ distorted polyhedra form a binuclear unit $(Fe_2O_{10})^{-14}$ in the edge and corner-sharing configuration. Different intrabilayer and interbilayer exchange pathways are proposed to induce various kinds of ME responses above and below the $T_N$.

In this work, the macroscopic and local magnetic properties are investigated by magnetization, heat capacity, and μSR measurement. The magnetic correlations observed in these measurements are corroborated from Raman spectroscopic data which reveals signatures of spin-phonon coupling below ~ 210 K.

## 2. Experimental techniques:

Magnetization measurements are carried out using the M/s. Quantum Design, USA make 9T Helium Re-liquefier based Vibrating sample magnetometer (VSM) attached with the Physical Property measurement system (PPMS). Sample in thin pellet form is used for specific heat measurement in M/s. Quantum Design made PPMS. The measurement is carried out in the temperature window 100 K≤T≤ 300 K for different fixed magnetic fields (0T, 1T, 5T) following relaxation methods. The muon spin relaxation (μSR) data has been collected at the ISIS pulsed muon facility, Rutherford Appleton Laboratory, United Kingdom. The temperature-dependent Raman measurements were performed in the backscattering geometry using M/s. LabRAM, HR Raman spectrometer attached to a Peltier-cooled charge-coupled device (CCD) detector. A laser source of wavelength 532 nm was used for excitation, and the sample temperature was varied between 80 – 400 K using a Linkam heating stage.

## 3. Results and Discussions:

## 3.1 Magnetization Study:

Figure 1(a) shows the temperature-dependent magnetization where a broad maximum is observed above the sharp AFM transition at 210 K. The magnetic ordering in the material is driven by two superexchange (SE) interactions. One is double oxygen-mediated SE (Fe-$O_1$-Fe) $J_1$ in between two Fe along the *c*-axis (figure 1(b)) having a SE angle of ~100°. Another one is $J_2$ resulting from the correlation between Fe along the *c*-axis and the Fe in the body center of the subcompartment (figure 1(b)) of the trirutile structure via Fe-$O_2$-Fe bond angle ~ 128°. The broad maximum above $T_N$ is a common feature of the trirutile antiferromagnets[15,16]. Despite the exchange angle of ~ 100°, $J_1$ gives rise to AFM pairing between Fe in the quasi binuclear unit $(Fe_2O_{10})^{-14}$ due to the direct *d-d* dimeric interaction ($J_{dim}$) along with the *p-d* interaction. The molar susceptibility data is fitted (inset of figure 1(a)) with the dimeric interaction model (Eq. 1) containing two spins; $S_1=5/2$ and $S_2=5/2$; $S=S_1+S_2=5$,

$$\chi = \frac{ANg^2\mu_B}{3k_BT} \frac{\sum_{S=0}^{5} S(s+1)(2S+1)e^{-\frac{E}{k_BT}}}{\sum_{S=0}^{5}(2S+1)e^{-\frac{E}{k_BT}}} \quad [1]$$

All the terms in the equation have the usual meaning whereas, the pre-factor A represents the deviation from the ideal dimeric interaction as the two distorted iron polyhedra are sharing edges and the spin-spin interaction energy: $E=-J_{dim}S(S+1)$[24]. The fitting results the dimeric interaction strength as $J_{dim}/k_B$=35.49(0.05) and deviation parameter as A=0.7055(0.0008). The strength of the dimeric interaction depends on the hybridization of the *d*-orbitals with the ligands *p*-orbital in the edge shared $Fe^{3+}$ bi-nuclear unit. Orbital structure and the splitting of *d*-orbital in this type of distorted edge shared configuration is not like the common regular octahedra sharing edges with 90° sharing angles between magnetic ions bridged by O ligands. Small but finite deviation from $D_{4h}$ symmetry at $Fe^{3+}$ polyhedra is concluded in different structural analyses[22]. The dimeric interaction explains the broad feature in the susceptibility curve, suggesting the presence of a weak/short-ranged magnetic correlation above the long-range ordering at 210 K. The magnetic susceptibility below the sharp transition at 210 K shows a change in slope at ~150 K. This anomaly is not reported elsewhere for FTO. A rise in susceptibility is observed below 50 K. Both the signatures indicate the possibility of different magnetic correlations rather than static magnetic order. The presence of magnetic correlation above $T_N$, an anomaly at 150 K, and the rise in susceptibility are further inspected by specific heat and muon spin rotation (µSR) measurements.

## 3.2 Specific Heat measurement:

The measured total heat capacity ($C_P$) of FTO is subtracted by the specific heat of the non-magnetic isostructural $Ga_2TeO_6$ to obtain the magnetic specific heat ($C_{mag}$) from 100 -300 K. A sharp anomaly at ~204 K corresponding to the AFM transition is observed (figure 2) as shown in the temperature scaled $C_{mag}$. Apart from the sharp peak, a conspicuous change is observed in the curve at ~150 K followed by a bump (figure 2), consistent with the sudden change in slope sudden slope change in magnetic susceptibility (figure 1(a)) as discussed. The magnetic entropy ($S_{mag}$) is calculated by integrating the $C_{mag}/T$ over the temperature 100 K to 300 K. $S_{mag}$ is normalized by the saturation value $S_{mag}= Rln(2S+1)$ with spin value $S=5/2$ (figure 2). The normalized entropy shows that only ~50% entropy is released below $T_N$. Entropy value attains ~62% to 14.9 ($Rln6$) on heating above $T_N$ for S=5/2 up to 300 K. This residual magnetic entropy implies the presence of magnetic correlation above the $T_N$ as seen in the broad maximum in magnetization data [18]. Magnetic field-dependent specific heat scaled by temperature ($C_P/T$) is shown in the inset of figure 2. No significant change is observed in $C_P$ on the application of the magnetic field. Thus, AFM transition ($T_N$), magnetic correlation above $T_N$, and the anomaly at 150 K can be revalidated by specific heat measurement. The magnetic transition and dynamics are further studied by µSR, as discussed in the next section.

## 3.4 Muon Spin Relaxation Study:

The local magnetic properties and spin dynamics are probed by zero fields (ZF) and longitudinal field (LF) µSR spectroscopy to validate the magnetic signatures. Temperature-dependent ZF (figure 3(a)) and LF (figure 4(a, b)) µSR spectra are fitted with stretched exponential function with a fixed background ($Bkg$) $A = A_0 e^{-(\lambda t)^\beta} + Bkg$. The temperature variation of the obtained initial asymmetry parameter ($A$), muon spin relaxation parameter ($\lambda$), and stretched exponential parameter ($\beta$) are shown in figure 3(b, c, and d, respectively). In general, the Stretch-exponential function can represent a wide range of magnetic phenomenon; like spin glass, magnetic frustration, and various kinds of exotic magnetic behavior in low-dimensional magnetic systems[25]. The dynamical characteristic features observed in the muon depolarization distinguish the type of magnetism present in the material[19]. The sharp decrease in asymmetric parameter at ~210 K corresponds to the AFM transition, consistent with our magnetization and neutron diffraction studies. Above $T_N$, the gradual exponential decay of $\lambda$ signifies the presence of a weak/short-ranged magnetic correlation due to dimeric interactions as evident from unsaturated magnetic entropy and broad feature in the magnetization data [26]. The $\lambda$ parameter rises below 215 K with a shoulder at ~200 K as the dynamics get slowed down due to AFM long-range ordering. After that, a peak at 150 K followed by a sharp fall is observed as the temperature decreases further. The $\beta$ value also changes from ~0.6 ($\beta \leq 0.6$ corresponds to 3$d$ ordering) to ~0.3 ($\beta \leq 0.5$ corresponds to the low dimensional ordering or kind of magnetic correlation)[27]. This behavior intrigues to suspect the spin dynamics associated with the 3-$d$ Heisenberg-like AFM in its ordered state. Though all the magnetic anomalies are present in the temperature variation of the A and $\lambda$, certain ambiguities are refurbished to conclude about the dynamical state of the system below 150 K. The diverging nature of $\lambda$ value at 150 K, and the absence of any oscillatory spectra suggest that this material exhibits dynamic spin correlation at low temperatures[28]. ZF-µSR time spectra do not recover 1/3 tail which is expected in spin freezing-like behavior[29,30]. The $\beta$ values decrease continuously from ~1 (at 300 K) to ~0.3 (at 1.8 K). In the case of spin freezing, the $\beta$ attains a value ~1/3 at spin freezing temperature. Thus the system may have a dynamic magnetic correlation instead of the only static uniform magnetic or static magnetic disorder[28].

LF muon spectra are collected at 140 K and 1.8 K to distinguish the spin dynamics of the system below $T_N$. Fitted LF spectra are shown in figure 4(a, b). Under LF, the dipolar interaction of nuclear spin with muon spin gets decoupled. Electronic interaction dominantly contributes to muon spin relaxation. The relaxation gets quenched by the spin-locking effect if the internal magnetic field is of the order of the applied magnetic field. A systematic upward shift is observed with increasing magnetic field in both the temperature. This suggests the presence of a static internal field below $T_N$, though considerable unquenched relaxation is observed even up to 4000 Oe applied field in both the temperatures. The presence of considerable unquenched relaxation with a longtime residual slope of µSR spectra signifies the coexistence of a dynamically fluctuating field and a static internal magnetic field. The Neutron diffraction suggests the presence of long-range AFM order (static) at low temperature as the magnetic Bragg's peaks are observed below $T_N$. The elastic neutron scattering does not sense the dynamical characteristics of the spins as it only takes the snapshot of the long-range ordered moments. The unquenched relaxation is higher at 140 K than 1.8 K for all the applied magnetic fields. Magnetic field variations of $\lambda_{LF}$ at 140 K and 1.8 K are plotted in figure 4(c). A narrow peak is observed at ~ 500 Oe, which implies the competition of static and dynamic components of the local field at the muon site. These two components cannot be segregated due to the low time resolution of the pulsed muon beam at ISIS. Such a peak suggests the existence of a broad distribution of spin-spin correlation. The field variation of $\lambda_{LF}$ has no such peak at 1.8 K. The residual slope of the unquenched spectra, and the field variation of $\lambda_{LF}$ confirm the presence of dynamic spin correlation that persists at 1.8 K.

In the layered structure of $Fe^{3+}$, interbilayer and intrabilayer competing exchange interactions may drive such a dynamical magnetic behavior. The time resolution of the pulse muon at ISIS is not sufficient to understand the three expected magnetic phases in the three temperature regimes (color shaded area of figure 3(c)), *i.e.* (I) $T > T_N$, (II) $T_N > T > 150$ K, and (III) T<150 K. Continuous muon beam with better time resolution may help to understand the dynamics of the spins at low temperature[31]. However, this is beyond the scope of the present study. In the next section, the spin-phonon correlation by Raman spectroscopy with an understanding of magnetism in the temperature regime (I) and (II) will be discussed.

### 3.5 Signatures of Spin-Phonon Coupling by Raman spectroscopic study:

Raman spectroscopy technique is employed to investigate the phonon behavior with temperature and the possible effects of magnetic ordering. According to the factor group analysis, the Raman active modes for an inverse trirutile tetragonal phase include $E_g$, $A_{1g}$, $B_{1g}$, and $B_{2g}$ symmetries[32]. Table I enlists the phonon frequencies, symmetries, and the atoms involved in the vibrations for all the Raman active modes based on the report by Hauseler [32]. Figure 5(a) shows the Raman spectrum at room temperature fitted with Lorentzian function where modes are labeled as *P1* to *P10*. Raman spectra at a few typical temperatures are shown in figure 5(b). Most of the phonons show a redshift with increasing temperature (shown by dotted lines for P1, P3, and P4) owing to the phonon anharmonicity. On the contrary, it is interesting to note that some of the modes display an anomalous behavior in frequency below $T_N$. A color map representing the shift in frequency with temperature for the modes *P1* and *P2* is shown in the figure 5(c). It can be observed that *P2* deviates considerably while *P1* hardly shows a deviation from the usual cubic-anharmonic behavior. The temperature dependence of the frequency for a phonon mode can be expressed as[33,34].

$$\omega(T) = \omega(0) + \Delta\omega_{el-ph}(T) + \Delta\omega_{anh}(T) + \Delta\omega_{sp-ph}(T) \quad [2]$$

where, $\omega(0)$ denotes the phonon frequencies at 0K, whereas $\Delta\omega_{el-ph}$, $\Delta\omega_{anh}$, and $\Delta\omega_{sp-ph}$ respectively, represent the contribution to the frequency from electron-phonon interaction, phonon anharmonicity, and spin-phonon interaction. Considering the contribution of the individual term for FTO, the insulating nature accounts for $\Delta\omega_{el-ph}$ being negligible. $\Delta\omega_{anh}$ corresponds to the contribution arising from phonon-phonon anharmonic interactions. Considering a three-phonon anharmonic process, the temperature dependence of frequency can be expressed as,

$$\omega_{anh}(T) = \omega(0) + C\left[1 + \frac{2}{e^{\frac{\hbar\omega}{2kT}}-1}\right] \quad [3]$$

where, $C$ and $\omega(0)$ are the coefficient of anharmonicity and frequency at absolute zero temperature, respectively. Figure 6 shows the temperature-dependent frequencies for the modes *P1-P10* fitted with the cubic-anharmonicity (Eq. 3). On careful observation, it can be seen that most of the modes exhibit a significant deviation from the anharmonic trend (shown by solid lines) below 210 K, which may be attributed to spin-phonon coupling. Signatures of spin-phonon coupling can also be realized by the temperature variation of the linewidth of individual modes. Linewidth is correlated with the phonon lifetime, which can be modified due to spin-phonon coupling. Figure 7 shows the temperature variation of the linewidth of different modes fitted with cubic-anharmonicity. Where the modes *P2*, *P4*, *P6*, *P7* show a slight deviation from cubic anharmonic behavior below 210K. The deviation of linewidth signifies the modulation of phonon lifetime due to spin-phonon coupling[35,36]. The magnetic ordering below ~ 210 K may promote a coupling between the spin and lattice (phonon) degrees of freedom, thus giving rise to an anomalous deviation in frequency and linewidth of specific modes from the expected anharmonic trend (figure. 6 and 7). Within the mean-field approximation, the deviation in frequency can be written as [37].

$$\Delta\omega_{sp-ph}(T) = \omega_{exp}(T) - \omega_{anh}(T) = \lambda_{sp}\langle S_i \cdot S_j \rangle \quad [4]$$

where, $\omega_{exp}(T)$ is the frequency in the absence of spin-phonon coupling. The bracketed quantity refers to the spin-spin correlation (SSC) between the adjacent magnetic sites i and j located on the opposite sublattices, and $\lambda_{sp}$ is the spin-phonon coupling constant. The strength of spin-phonon coupling can be estimated by evaluating the SSC function, which is defined as $\langle S_i.S_j \rangle = -S^2 \Phi (T)$ in terms of short-range order parameter $\Phi (T)$ and spin $S$[38]. The magnitude of the correlation function decreases rapidly on approaching $T_N$ and becomes zero above it. In order to take into account the correlation effects above $T_N$, $\Phi (T)$ is calculated following the two-spin cluster approach by Cottom *et al.* for different spin values ($S$=1/2, 1, 2, and 5/2)[38,39]. Since the spin-state of $Fe^{3+}$ is approximately 5/2, we estimate $\lambda_{sp}$ in the present case using the $\Phi$ values for $S$=5/2 in the following equation[37,40].

$$\lambda_{sp} = - \frac{\Delta \omega_{sp-ph}(T_{low})}{[\Phi(T_{low}) - \Phi(2T_N)]S^2} \qquad [5]$$

where the deviation $\Delta \omega_{sp-ph}(T_{low})$ and $\Phi(T_{low})$ are considered at 80 K (lowest temperature in our case). The values of the spin-phonon coupling constant for the modes *P2*, *P6*, *P7*, and *P10*, along with the other modes, are listed in Table I. It is found that $\lambda_{sp}$ has the highest value of *1.2 cm$^{-1}$* for *P6* mode, while *P2* shows a value of *0.5 cm$^{-1}$*.

To validate the spin-phonon coupling in the magnetically ordered phase of FTO, a detailed study of exchange pathways has been done that strongly depends on the bond angles and bond lengths involving the magnetic ions. The AFM nature of FTO can be understood based on semi-empirical Goodenough-Kanamori (GK) rules because it possesses two SE pathways with Fe-O1-Fe ($J_1$) bond angle being nearly equal to 100° while the Fe-O2-Fe ($J_2$) angle is approximately 128°. From the GK rules, it can be inferred that $J_2$ will strongly favor antiferromagnetism in the ordered phase of FTO[41,42]. In the case of $J_1$ interaction pathway, the Fe-O1 bond lengths are equal that results in the symmetric vibrations of the atoms. However, the Fe-O2 bond lengths of the Fe-O2-Fe unit ($J_2$, SE pathway) are unequal and hence, would give rise to asymmetric vibrations of phonon modes. Therefore, we propose that SE interaction ($J_2$), mediated via asymmetric interatomic bond distances is responsible for phonon renormalization due to spin-spin correlation. The SE/spin-phonon coupling pathway may also be corroborated with our EXAFS results shown elsewhere [22]. The EXAFS analysis shows a significant value of the Debye-Waller factor (0.0073(2)) and flexibility to have spin-lattice coupling in that pathway. The corner-sharing of the *c*-axis and the sub-compartmental Fe polyhedra play extra constrain during magnetic ordering. So the ME coupling below $T_N$ is mediated by the spin-lattice coupling via $J_2$.

## 4. Conclusion:

We conclude that the intrabilayer exchange ($J_1+J_{dim}$) and interbilayer exchange ($J_2$) interaction induce distinguishable magnetic features in the three separate temperature regimes; (i) $T >T_N$, (ii) $T_N>T>150$ K, and (iii) $T<150$ K. The broad anomaly at $T>T_N$ is explained by the 5/2-5/2 dimeric (*d-d*) interaction. Another anomaly at 150 K (T < $T_N$), which was not studied previously, is observed separately in magnetization, heat capacity in µSR measurement. Signatures of distinct features of spin dynamics are revealed by µSR measurement in the three temperature regimes. The observed features are intriguing to reinvestigate the spin dynamics with continuous muon beam, inelastic neutron, and polarized neutron to have a more resolved picture of the magnetic phase diagram.

Detailed investigation of frequency shift and linewidth of Raman spectra reveals renormalization of phonon spectra below $T_N$. The renormalization is attributed to the spin-phonon coupling mediated by the Fe-O2-Fe pathway. Spin-phonon coupling is believed to play dominating role in the ME behavior below $T_N$. The Raman spectroscopic study also confirms that the $Fe^{3+}$ solely drives the SPC and magnetic order in the system. We quantified the spin-phonon coupling strength for various modes that are in the range of 0.1-1.2 cm$^{-1}$ by following two spin cluster methods. AFM with reasonable SPC coupling is of great technological importance, and we believe FTO may be a good candidate for such applications.

**Acknowledgment:** We acknowledge UGC-DAE-CSR Mumbai, India, for the project grant (Sanction No. CRS-M-187, 225).

References

[1]     Khomskii D I and Sawatzky G A 1997 Interplay between spin, charge and orbital degrees of freedom in magnetic oxides *Solid State Commun.* **102** 87

[2]     Orenstein J and Millis A J 2000 Advances in the physics of high-temperature superconductivity *Science* **288** 468

[3]     Ideue T, Kurumaji T, Ishiwata S and Tokura Y 2017 Giant thermal Hall effect in multiferroics *Nat. Mater.* **16** 797

[4]     Tokura Y and Nagaosa N 2000 Orbital physics in transition-metal oxides *Science* **288** 462

[5]     Lee J H, Fang L, Vlahos E, Ke X, Jung Y W, Kourkoutis L F, Kim J W, Ryan P J, Heeg T, Roeckerath M, Goian V, Bernhagen M, Uecker R, Hammel P C, Rabe K M, Kamba S, Schubert J, Freeland J W, Muller D A, Fennie C J, Schiffer P, Gopalan V, Johnston-Halperin E and Schlom D G 2010 A strong ferroelectric ferromagnet created by means of spin-lattice coupling *Nature* **466** 954

[6]     Eerenstein W, Mathur N D and Scott J F 2006 Multiferroic and magnetoelectric materials *Nature* **442** 759

[7]     Mochizuki M, Furukawa N and Nagaosa N 2011 Theory of spin-phonon coupling in multiferroic manganese perovskites $RMnO_3$ *Phys. Rev. B* **84** 144409

[8]     Ramesh R and Spaldin N A 2007 Multiferroics: Progress and prospects in thin films *Nat. Mater.* **6** 21

[9]     Moya X and Mathur N D 2017 Thermal hall effect: Turn your phonon *Nat. Mater.* **16** 784

[10]    Calero C, Chudnovsky E M and Garanin D A 2005 Field dependence of the electron spin relaxation in quantum dots *Phys. Rev. Lett.* **95** 166603

[11]    Son J, Park B C, Kim C H, Cho H, Kim S Y, Sandilands L J, Sohn C, Park J G, Moon S J and Noh T W 2019 Unconventional spin-phonon coupling via the Dzyaloshinskii–Moriya interaction *npj Quantum Mater.* **4** 17

[12]   Kunnmann W, La Placa S, Corliss L M, Hastings J M and Banks E 1968 Magnetic structures of the ordered trirutiles $Cr_2WO_6$, $Cr_2TeO_6$ and $Fe_2TeO_6$ *J. Phys. Chem. Solids* **29** 1359

[13]   Hornreich R M 1969 The magnetoelectric effect: Some likely candidates *Solid State Commun.* **7** 1081

[14]   Buksphan S, Fischer E and Hornreich R M 1972 Magnetoelectric and Mössbauer studies of Fe2 TeO6 *Solid State Commun.* **10** 657

[15]   Dehn J T, Newnham R E and Mulay L N 1968 Mixed magnetic ordering: Magnetic susceptibility and mossbauer studies on iron(IH) tellurate (Fe2TeO6) *J. Chem. Phys.* **49** 3201

[16]   Yamaguchi M and Ishikawa M 1994 Magnetic Phase Transitions in Inverse Trirutile-Type Compounds *J. Phys. Soc. Japan* **63** 1666

[17]   Zhu M, Matsumoto M, Stone M B, Dun Z L, Zhou H D, Hong T, Zou T, Mahanti S D and Ke X 2019 Amplitude modes in three-dimensional spin dimers away from quantum critical point *Phys. Rev. Res.* **1** 033111

[18]   Zhu M, Do D, Dela Cruz C R, Dun Z, Cheng J G, Goto H, Uwatoko Y, Zou T, Zhou H D, Mahanti S D and Ke X 2015 Ferromagnetic superexchange in insulating C r2Mo O6 by controlling orbital hybridization *Phys. Rev. B* **92** 094419

[19]   Dalmas De Réotier P and Yaouanc A 1997 Muon spin rotation and relaxation in magnetic materials *J. Phys. Condens. Matter* **9** 9113

[20]   Pal P, Abdullah M F, Sahoo A, Kuila S, Chandrakanta K, Jena R, Vishwakarma P N, Kaushik S D and Singh A K 2019 Doping induced modification in magnetism and magnetoelectric coupling at room temperature in Fe2Te(1-x)NbxO6 *Phys. B Condens. Matter* **571** 193

[21]   Pal P, Sahoo A, Abdullah M F, Kaushik S D, Vishwakarma P N and Singh A K 2018 Substantial magnetoelectric coupling in nanocrystalline-Fe2TeO6 at room temperature *J. Appl. Phys.* **124** 164110

[22] Pal P, Kaushik S D, Kuila S, Rajput P, Vishwakarma1 P N and Singh A K 2020 $d^5$-off-centering induced ferroelectric and magnetoelectric correlations in trirutile-$Fe_2TeO_6$ arXiv:2011.08017

[23] Khomskii D I, Kugel K I, Sboychakov A O and Streltsov S V. 2016 Role of local geometry in the spin and orbital structure of transition metal compounds *J. Exp. Theor. Phys.* **122** 484

[24] Khuntia P and Mahajan A V. 2010 Magnetic susceptibility and heat capacity of a novel antiferromagnet: $LiNi_2P_3O_{10}$ and the effect of doping *J. Phys. Condens. Matter* **22** 296002

[25] Phillips J C 1996 Stretched exponential relaxation in molecular and electronic glasses *Reports Prog. Phys.* **59** 1133

[26] Steer C A, Blundell S J, Coldea A I, Marshall I M, Lancaster T, Battle P D, Gallon D, Fargus A J and Rosseinsky M J 2003 A μSR study of the spin dynamics in Ir-diluted layered manganites *Phys. B Condens. Matter* **326** 513

[27] Kawasaki Y, Gavilano J L, Keller L, Schefer J, Christensen N B, Amato A, Ohno T, Kishimoto Y, He Z, Ueda Y and Itoh M 2011 Magnetic structure and spin dynamics of the quasi-one-dimensional spin-chain antiferromagnet $BaCo_2V_2O_8$ *Phys. Rev. B - Condens. Matter Mater. Phys.* **83** 064421

[28] Li Y, Adroja D, Biswas P K, Baker P J, Zhang Q, Liu J, Tsirlin A A, Gegenwart P and Zhang Q 2016 Muon Spin Relaxation Evidence for the U(1) Quantum Spin-Liquid Ground State in the Triangular Antiferromagnet $YbMgGaO_4$ *Phys. Rev. Lett.* **117** 097201

[29] Uemura Y J, Yamazaki T, Harshman D R, Senba M and Ansaldo E J 1985 Muon-spin relaxation in AuFe and CuMn spin glasses *Phys. Rev. B* **31** 546

[30] Uemura Y J, Keren A, Kojima K, Le L P, Luke G M, Wu W D, Ajiro Y, Asano T, Kuriyama Y, Mekata M, Kikuchi H and Kakurai K 1994 Spin fluctuations in frustrated kagomé lattice system $SrCr_8Ga_4O_{19}$ studied by Muon spin relaxation *Phys. Rev. Lett.* **73** 3306

[31] Lancaster T, Giblin S R, Allodi G, Bordignon S, Mazzani M, De Renzi R, Freeman P G,

Baker P J, Pratt F L, Babkevich P, Blundell S J, Boothroyd A T, Möller J S and Prabhakaran D 2014 Stripe disorder and dynamics in the hole-doped antiferromagnetic insulator La$_{5/3}$Sr$_{1/3}$CoO$_4$ *Phys. Rev. B.* **89** 020405

[32] Haeuseler H 1981 Infrared and Raman spectra and normal coordinate calculations on trirutile-type compounds *Spectrochim. Acta Part A Mol. Spectrosc.* **37** 487

[33] Balkanski M, Wallis R F and Haro E 1983 Anharmonic effects in light scattering due to optical phonons in silicon *Phys. Rev. B* **28** 1928

[34] Saha S, Muthu D V S, Singh S, Dkhil B, Suryanarayanan R, Dhalenne G, Poswal H K, Karmakar S, Sharma S M, Revcolevschi A and Sood A K 2009 Low-temperature and high-pressure Raman and x-ray studies of pyrochlore Tb2 Ti2 O7: Phonon anomalies and possible phase transition *Phys. Rev. B* **79** 064109

[35] Rawat R, Phase D M and Choudhary R J 2017 Spin-phonon coupling in hexagonal Sr$_{0.6}$Ba$_{0.4}$MnO$_3$ *J. Magn. Magn. Mater.* **441** 398

[36] Singh B, Vogl M, Wurmehl S, Aswartham S, Büchner B and Kumar P 2020 Coupling of lattice, spin, and intraconfigurational excitations of Eu 3+ in Eu 2 ZnIr O 6 *Phys. Rev. Res.* **2** 043179

[37] Aytan E, Debnath B, Kargar F, Barlas Y, Lacerda M M, Li J X, Lake R K, Shi J and Balandin A A 2017 Spin-phonon coupling in antiferromagnetic nickel oxide *Appl. Phys. Lett.* **111** 252402

[38] Lockwood D J and Cottam M G 1988 The spin-phonon interaction in FeF2 and MnF2 studied by Raman spectroscopy *J. Appl. Phys.* **64** 5876

[39] Cottam M G and Lockwood D J 2019 Spin-phonon interaction in transition-metal difluoride antiferromagnets: Theory and experiment *Low Temp. Phys.* **45** 78

[40] Poojitha B, Rathore A, Kumar A and Saha S 2020 Signatures of magnetostriction and spin-phonon coupling in magnetoelectric hexagonal 15R-BaMnO$_3$ *Phys. Rev. B* **102** 134436

[41]   Goodenough J B 1955 Theory of the role of covalence in the perovskite-type manganites [La,M(II)]MnO$_3$ *Phys. Rev.* **100** 564

[42]   Kanamori J 1959 Superexchange interaction and symmetry properties of electron orbitals *J. Phys. Chem. Solids* **10** 87

**Figure Captions:**

**Figure 1.** (a) Temperature variation of magnetic susceptibility ($\chi$) in zero-field-cooled (ZFC) and field-cooled (FC) mode. The inset shows *Fe-Fe* dimeric interaction fit (red solid curve) of molar susceptibility considering 5/2 spin state. The marked section in the inset shows the change in slope in molar susceptibility ($\chi_{molar}$ in ZFC mode). (b) Schematic diagram of the unit cell along with different exchange pathways responsible for magnetic ordering.

**Figure 2.** Temperature scaled magnetic heat capacity of FTO, and magnetic entropy normalized considering 5/2 (blue). Inset: Magnetic field (0T, 1T, and 5T) dependent heat capacity.

**Figure 3.** (a) The zero-field (ZF) µSR spectra at selected temperatures. The solid lines show the stretched exponential fitting $\left(A = A_0 e^{-(\lambda t)^\beta}\right)$. (b) Temperature dependency of asymmetry ($A_0$). (c) Temperature-dependent relaxation parameter ($\lambda$). Inset: exponential variation of spin relaxation parameter $\lambda$ above $T_N$. (d) Temperature dependency of stretched exponent $\beta$.

**Figure 4.** The LF muon spectra at the different magnetic fields at (a) 140 K and (b) 1.8 K temperatures. The solid lines show the fitting to the data. (c) Magnetic (LF) field variation of $\lambda_{LF}$ at 140 K and 1.8 K below. Dotted lines are for a better line of sight.

**Figure 5.** (a) Raman spectrum of FTO at 300 K fitted with Lorentzian function (b) Raman spectra at a few fixed temperatures (c) Color map of modes *P1* and *P2* showing a shift in frequency with the temperature where *P2* exhibits anomaly below ~ 210 K.

**Figure 6.** Temperature dependence of frequency and cubic anharmonic fit (solid line) for the modes *P1 - P10*.

Figure 7. Temperature dependence of line width and cubic anharmonic fit (solid line) for the modes *P1 - P10*.

TABLES:

TABLE I: *Phonon modes along with the symmetries and atoms involved in the vibration. Spin-phonon coupling constants ($\lambda_{SP}$) for all the modes in S=5/2 spin state of $Fe^{3+}$.*

| Mode | Phonon Frequency (cm$^{-1}$) | Symmetry | Atoms Involved | $\lambda_{sp}$ (cm$^{-1}$) |
|---|---|---|---|---|
| P1 | 295 | $A_{1g}/B_{2g}$ | Fe /Fe, O | 0.1 |
| P2 | 314 | $E_g$ | Fe, O | 0.5 |
| P3 | 419 | $E_g$ | Fe, O | 0.1 |
| P4 | 463 | $E_g$ | Fe, O | 0.2 |
| P5 | 562 | $B_{1g}$ | O | 0.5 |
| P6 | 642 | $E_g$ | Fe, O | 1.2 |
| P7 | 649 | $B_{2g}/A_{1g}$ | O | 0.6 |
| P8 | 752 | $E_g/A_{1g}$ | Fe, O/ O | 0.5 |
| P9 | 760 | $B_{2g}$ | Fe, O | 0.6 |
| P10 | 785 | $B_{2g}/A_{1g}$ | O/ Fe, O | 0.9 |

Figure 1.

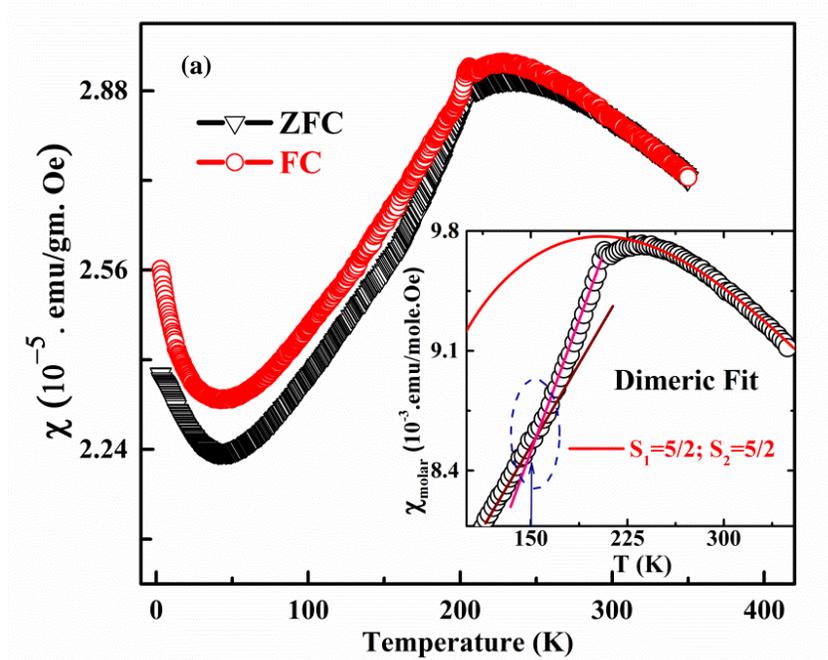

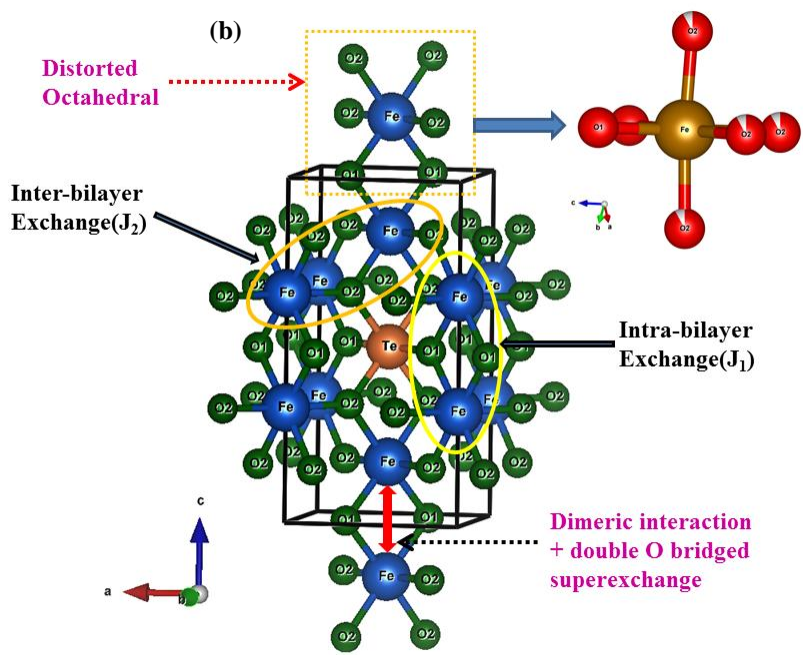

Figure 2.

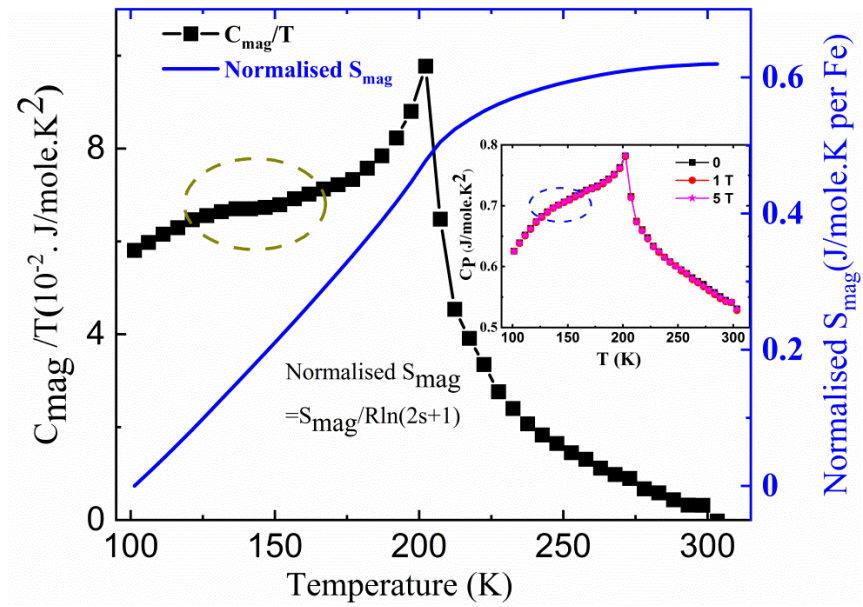

Figure 3.

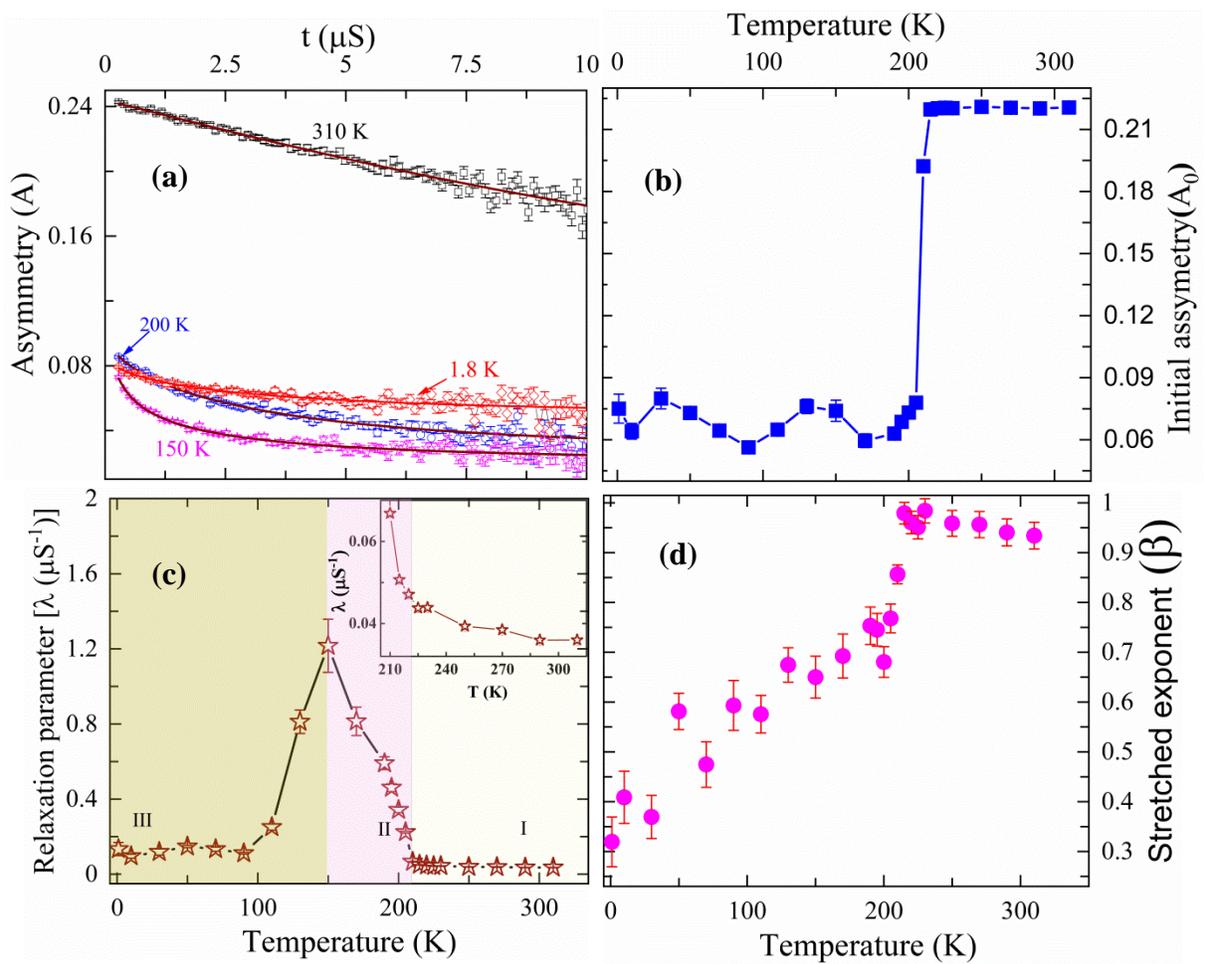

Figure 4.

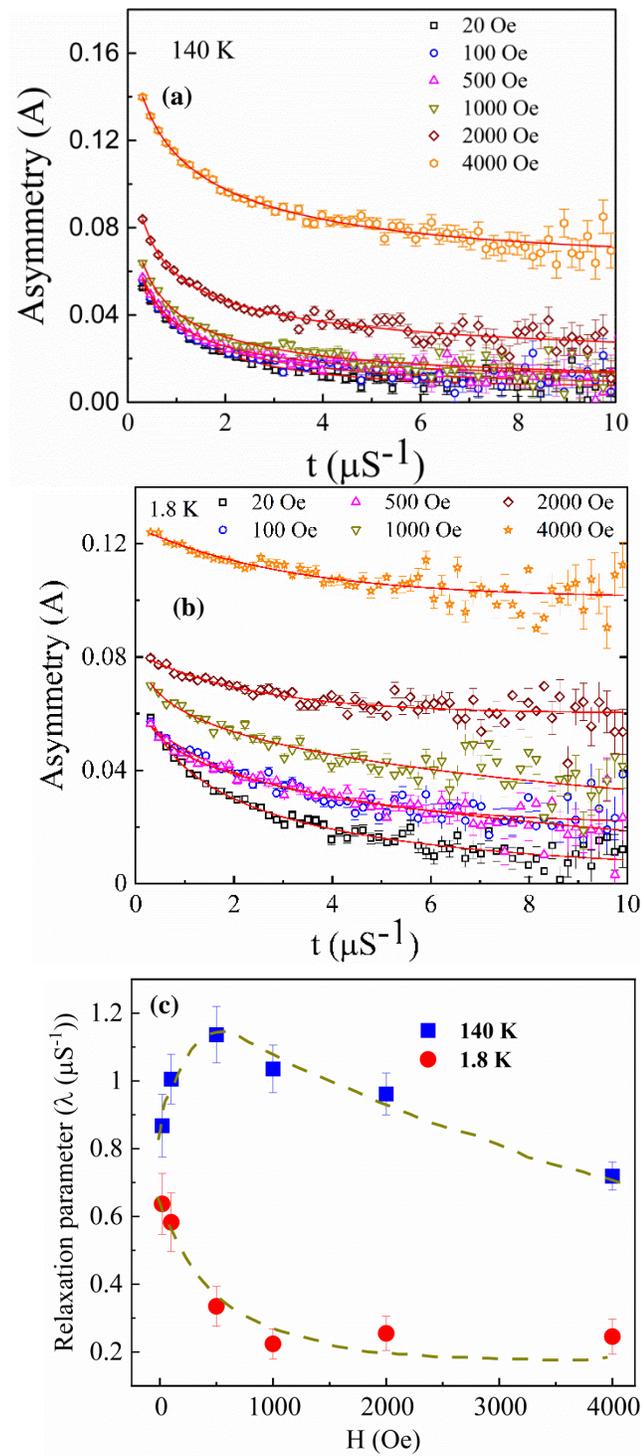

Figure 5.

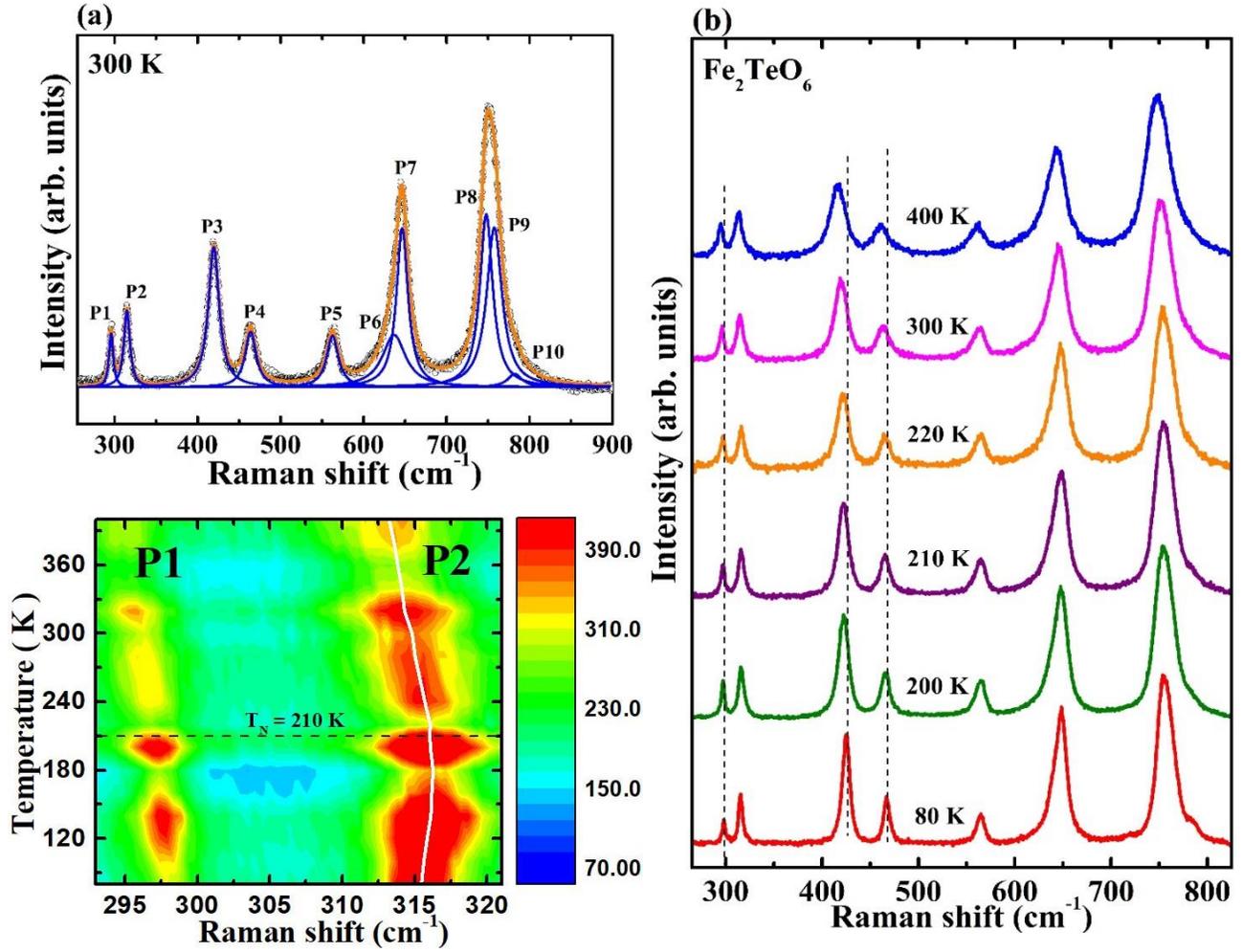

Figure 6.

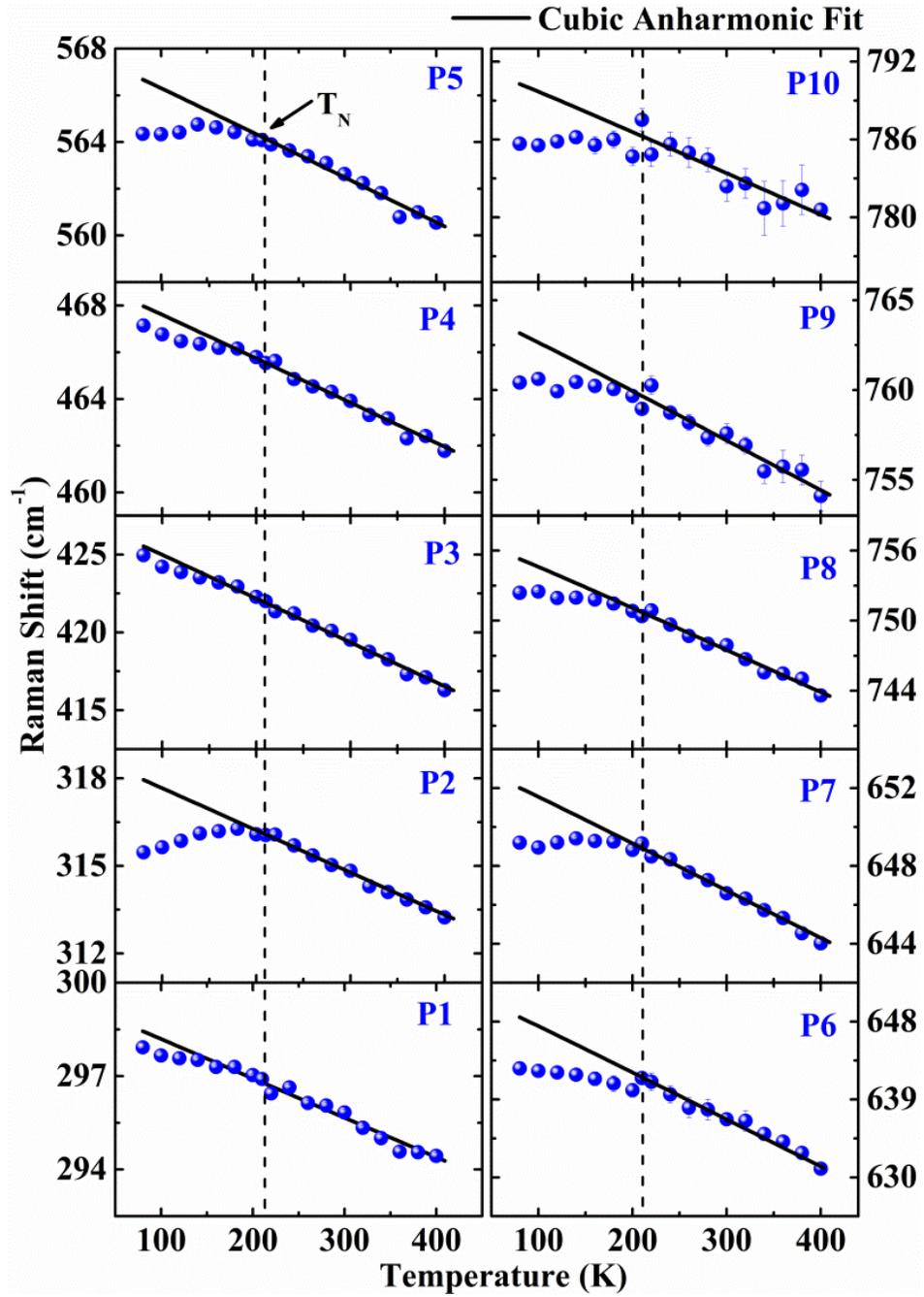

Figure 7.

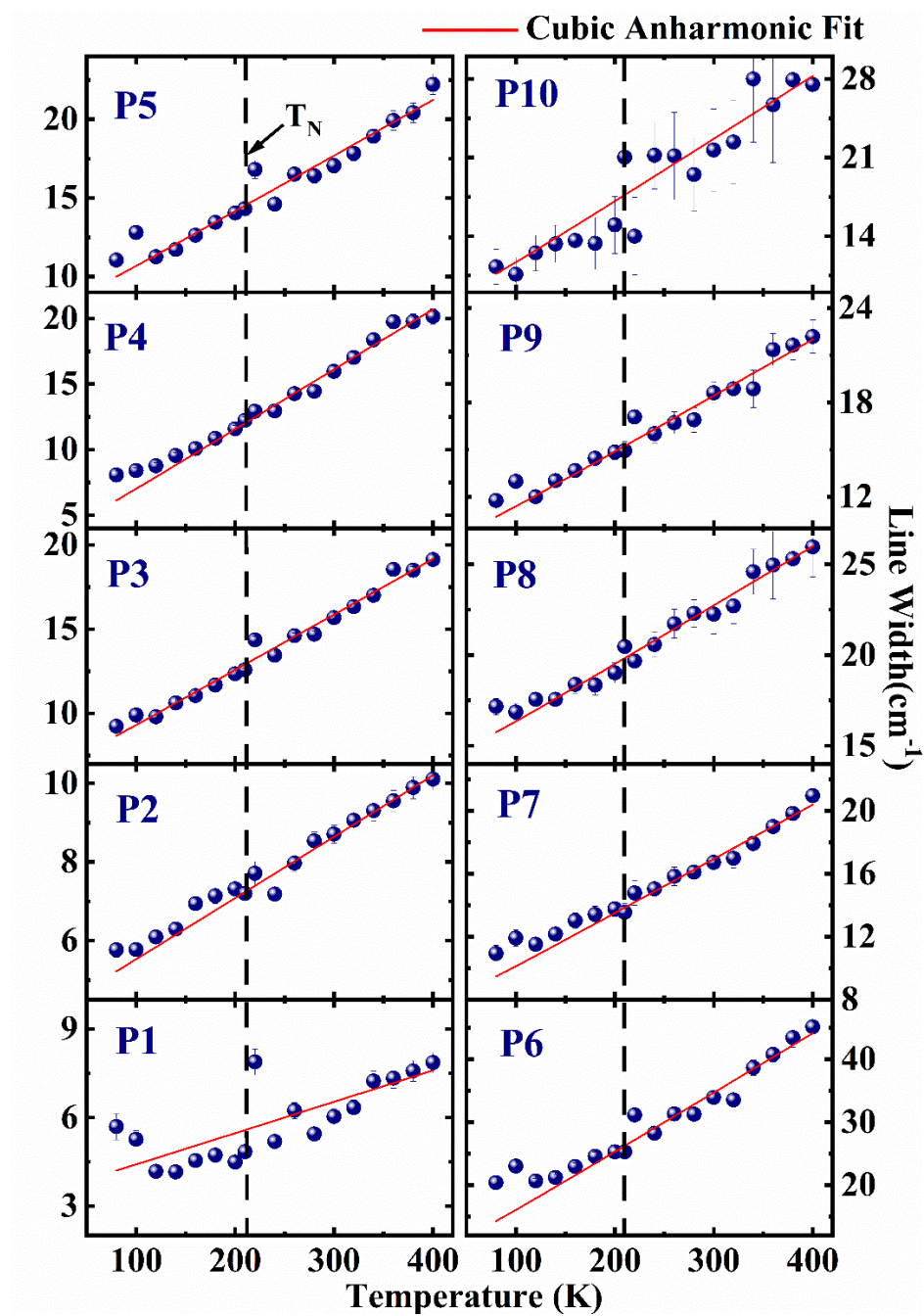